\documentclass[twocolumn,twoside,superscriptaddress,prl,aps]{revtex4}
\usepackage{amsmath,mathrsfs,amsbsy,color,graphicx,bm,amsthm,amsfonts}
\usepackage{units}
\usepackage{bbm}
\usepackage{color}
\usepackage[colorlinks,citecolor=blue]{hyperref}

\newcommand{\ket}[1]{|#1\rangle}
\newcommand{\bra}[1]{\langle #1|}

\usepackage{braket}

\newcommand{\mr}[1]{\mathrm{#1}}

\begin{document}
\title{Calculating the Green's function of two-site Fermionic Hubbard model in a photonic system}
\author{Jie Zhu}
\affiliation{Key Laboratory of Quantum Information, University of Science and Technology of China, CAS, Hefei, 230026, China}
\affiliation{CAS Center for Excellence in Quantum Information and Quantum Physics, Hefei, 230026, China}
\author{Yuya O. Nakagawa}
\email{nakagawa@qunasys.com}
\affiliation{QunaSys Inc., Aqua Hakusan Building 9F, 1-13-7 Hakusan, Bunkyo, Tokyo 113-0001, Japan}
\author{Chuan-Feng Li}
\affiliation{Key Laboratory of Quantum Information, University of Science and Technology of China, CAS, Hefei, 230026, China}
\affiliation{CAS Center for Excellence in Quantum Information and Quantum Physics, Hefei, 230026, China}
\author{Guang-Can Guo}
\affiliation{Key Laboratory of Quantum Information, University of Science and Technology of China, CAS, Hefei, 230026, China}
\affiliation{CAS Center for Excellence in Quantum Information and Quantum Physics, Hefei, 230026, China}
\author{Yong-Sheng Zhang}
\email{yshzhang@ustc.edu.cn}
\affiliation{Key Laboratory of Quantum Information, University of Science and Technology of China, CAS, Hefei, 230026, China}
\affiliation{CAS Center for Excellence in Quantum Information and Quantum Physics, Hefei, 230026, China}

\begin{abstract}
The Green's function has been an indispensable tool to study many-body systems that remain one of the biggest challenges in modern quantum physics for decades. The complicated calculation of the Green's function impedes the research of many-body systems. The appearance of the noisy intermediate-scale quantum devices and quantum-classical hybrid algorithm inspire a new method to calculate the Green's function. 
Here we design a programmable quantum circuit for photons with utilizing the polarization and the path degrees of freedom to construct a highly-precise variational quantum state of a photon, and first report the experimental realization for calculating the Green's function of the two-site Fermionic Hubbard model, a prototypical model for strongly-correlated materials, in photonic systems.
We run the variational quantum eigensolver to obtain the ground state and excited states of the model, and then evaluate the transition amplitudes among the eigenstates. The experimental results present the spectral function of the Green's function, which agrees well with exact results.
Our demonstration provides the further possibility of the photonic system in quantum simulation and applications in solving complicated problems in many-body systems, biological science and so on.
\end{abstract}
\date{\today}

\maketitle

{\it Introduction.}
Strongly-correlated systems~\cite{Dagotto2015}, where interactions among particles are strong enough to alter properties of the systems from ones of non-interacting ones, exhibit a wide variety of interesting phenomena such as high-temperature superconductivity~\cite{Bednorz1986}.
The Green's function has been an essential theoretical tool to tackle such systems for many decades including serving as the basis for the diagrammatic calculation in high-energy physics and condensed matter physics~\cite{bonch2015green, abrikosov2012methods, fetter2012quantum} and so on.
It captures various properties of the system; especially in condensed matter physics, the dispersion relation of quasi-particle excitations of a system can be read from the Green's function, which gives insight for the nature of magnetic materials~\cite{coey2010magnetism} and topological insulators~\cite{hasan2010colloquium}.

Simulating quantum systems, including strongly-correlated systems, is one of the most promising applications of quantum computers to illustrate the practical computational speedup compared with classical computers.
However, the celebrated quantum algorithm~\cite{Kitaev1995,Cleve1998,shor1997}, such as phase estimation algorithm, Shor algorithm and so on, which solve specific problems much faster than classical computers, is not expected to be executable on such near-term quantum computers because it requires deep and complex quantum circuits that is impossible to actually realize without error correction. The idea inspired by variational computation provides the possibility to deal with some problems based on the near-term quantum devices without error correction, i.e.  the noisy intermediate-scale quantum devices (NISQ)~\cite{Preskill2018}. The quantum-classical hybrid algorithms~\cite{Peruzzo2014,cerezo2020variational,Endo2021review} based on the variational principle of quantum mechanics were proposed to utilize the NISQ for computing the  eigenenergies and eigenstates of quantum systems.
With the help of classical optimizers for quantum states and circuits, the NISQ is expected to show its advantages in solving difficult problems for classical computers.

Among various hardware-platforms for quantum simulations, the photonic system has huge potential due to the easy manipulate and the mature techniques~\cite{pieter2007,aspuru2012}.
The degrees of freedoms of photons can be used with high precision such as polarization \cite{poh2015}, path \cite{xiaominhu2020dim32}, orbital angular momentum \cite{oam10010}, and hybrid degrees of freedoms \cite{xilinwang2018}.
The manipulation of those degrees of freedoms can be performed by established techniques in quantum optics and it generates diverse quantum states in the Hilbert space defined by those degrees of freedoms. 

In this article, we present the first experimental evaluation of the Green's function of quantum many-body systems in photonic systems.
We employ a method proposed in Ref.~\cite{Endo2020calculation} to calculate the Green's function of a given quantum system on near-term quantum computers.
Unlike other methods to calculate the Green's function in the literature~\cite{Bauer2016, Kreula2016, Wecker2015, Kosugi2020, Pedernales2014, Roggero2019},
this method is based on the quantum-classical hybrid algorithm and has potential to be executed on the real quantum devices. 
We take the two-site Fermionic Hubbard model~\cite{Gutzwiller1963, Hubbard1963, Kanamori1963} as a primitive example of strongly-correlated systems and map it to the photonic system.
We obtain the ground state and the excited states by measuring the energy expectation values of the variational quantum states generated by programmable quantum circuits for photons.
The Green's function is computed based on the eigenenergies of those states and the transition amplitudes among them that are also evaluated by measurements in quantum circuits.
Our result illustrates the potential of photonic systems to solve quantum many-body systems, especially strongly correlated systems.

\begin{figure*}[!ht]
\includegraphics[width=18.3cm]{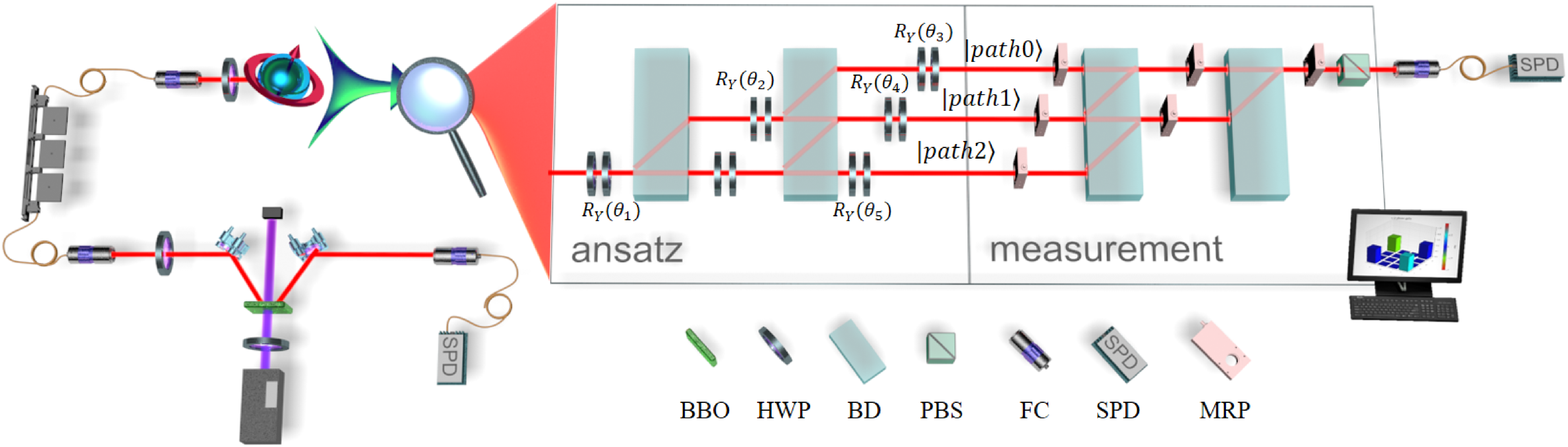}
\caption{\textbf{Experimental setup.} The heralded single-photon source is realized via the type-I phase-matching spontaneous parametric down conversion (SPDC) process in a joint $\beta$-barium-borate (BBO) crystal that is pumped by a 404 nm semiconductor laser. One photon is sent to the subsequent computing module via optical fiber and the other one is sent directly to the single photon detector (SPD) as a trigger. The first two beam displacers (BDs) and the operators $R_Y(\theta_i)$ prepare the ansatz states, where the operators $R_Y(\theta_i)$ are implemented with two half-wave plates (HWPs) fixed at $\theta/4$ and $0^\circ$ successively. Then the last two BDs and the motor rotating plates (MRPs) compose the measurement part. The MRPs are controlled by the computer to adjust the angles. The computer gives the updated $\theta_i$s according to the previous measurement result until the measurement result converging. BBO: $\beta$-barium-borate, HWP: half-wave plate, BD: beam displacer, PBS: polarization beam splitter, FC: fiber couple, SPD: single photon detector, MRP: motor rotating plate. }
\label{setupfigure}
\end{figure*}

{\it Model and notation.}
We consider the two-site Fermionic Hubbard model with particle-hole symmetry defined as
\begin{align} \label{eq: model}
 H_\mathrm{Hub} := -t \sum_{\sigma=\uparrow,\downarrow} \left( c_{0,\sigma}^\dagger c_{1,\sigma} + \rm{h.c.} \right) \nonumber \\
 + U \sum_{i=0,1}  n_{i,\uparrow} n_{i,\downarrow} 
  - \frac{U}{2} \sum_{i, \sigma}  n_{i,\sigma},
 \end{align}
where $c_{i, \sigma}, c_{i, \sigma}^\dag$ are creation and annihilation operators of electrons with spin $\sigma$ acting on site $i$, $n_{i,\sigma} = c_{i, \sigma}^\dagger c_{i, \sigma}$ is the number operator of electrons, and $t, U>0$ is a parameter for electron hopping and repulsion, respectively.
Since the model conserves the number of up-spin (down-spin) electrons $N_{\uparrow} = n_{0\uparrow} +n_{1\uparrow} ~(N_{\downarrow} = n_{0\downarrow} +n_{1\downarrow})$ simultaneously, we can diagonalize the Hamiltonian in each Hilbert space $\mathcal{H}_{(N_{\uparrow}, N_{\downarrow})}$ having $(N_{\uparrow}, N_{\downarrow})=(0,0),(0,1),\ldots,(2,2)$.
We introduce the spectral function of the Green's function at zero temperature for up-spin electrons in the Lehmann representation:  
\begin{align} \label{eq: spectral function}
A_{k\uparrow}(\omega) := - \frac{1}{\pi} \mathrm{Im} \left( \sum_n \frac{ |\braket{E_n^{(p)} |\tilde{c}_{k\uparrow}^\dag|\mathrm{GS}}|^2 }{\omega+E_\mathrm{GS}-E_n^{(p)}+i\eta}  \right. \nonumber \\
+ \left. \sum_m \frac{ |\braket{E_m^{(h)}|\tilde{c}_{k\uparrow}|\mathrm{GS}}|^2 }{\omega-E_\mathrm{G}+E_n^{(h)}+i\eta} \right),
\end{align}
where $k=\{0,~\pi\}$ is a momentum, $\tilde{c}_{k,\uparrow} := (c_{0,\uparrow}+(-1)^{k/\pi} c_{1,\uparrow})/\sqrt{2}$ is the momentum representation of the electron annihilation operators, $\ket{GS} (E_\mathrm{GS})$ is the ground state (energy), $\ket{E_n^{(h/p)}} (E_n^{(h/p)})$ is the eigenstate (eigenenergy) in the particle/hole sector (defined below), and $\eta>0$ is a small number determining the broadening of the spectral function.
The first (second) term in the right hand side is called the particle (hole) part.
We take the unit of $\hbar=1$ in this letter.
The spectral function is related with the imaginary part of the electron Green's function, and calculating the spectral function suffices to obtain the whole Green's function by the virtue of the Kramers-Kronig relation~\cite{bonch2015green,abrikosov2012methods, fetter2012quantum}.

Because the ground state of the model~\eqref{eq: model} resides in the Hilbert space $\mathcal{H}_{(1,1)}$, we choose the particle (hole) sector as $\mathcal{H}_{(2,1)} (\mathcal{H}_{(0,1)})$.
We choose the basis for $\mathcal{H}_{(1,1)}, \mathcal{H}_{(2,1)}$ and $\mathcal{H}_{(0,1)}$ as $\{ c_{0,\uparrow}^\dag c_{0,\downarrow}^\dag \ket{\mathrm{vac}},  c_{0,\uparrow}^\dag c_{1,\downarrow}^\dag \ket{\mathrm{vac}}, c_{0,\downarrow}^\dag c_{1,\uparrow}^\dag  \ket{\mathrm{vac}}, c_{1,\uparrow}^\dag c_{1,\downarrow}^\dag \ket{\mathrm{vac}} \}$,
$\{ c_{0,\uparrow}^\dag  c_{1,\uparrow}^\dag c_{1,\downarrow}^\dag \ket{\mathrm{vac}}, c_{0,\uparrow}^\dag  c_{0,\downarrow}^\dag c_{1,\uparrow}^\dag \ket{\mathrm{vac}} \}$,
and $\{ c_{0,\uparrow}^\dag  \ket{\mathrm{vac}}, c_{1,\uparrow}^\dag \ket{\mathrm{vac}} \}$, respectively, where $\ket{\mr{vac}}$ is a vacuum for electrons.
Since $\dim\mathcal{H}_{(1,1)}=4$ and $\dim\mathcal{H}_{(2,1)}=\dim\mathcal{H}_{(0,1)}=2$,
it is enough to consider a six-dimensional Hamiltonian on $\mathcal{H}_{(1,1)} \cup \mathcal{H}_{(2,1)}$,
\begin{equation} \label{eq: six-dim Ham}
H_\mr{six} = 
 \begin{pmatrix}
  0 & -t &  t  & 0 & 0 & 0\\
 -t &  -U & 0 & -t & 0 & 0\\
  t &  0 & -U & t & 0 & 0\\
 0 &  -t & t & 0 & 0 & 0 \\ 
 0 &  0 & 0 & 0 & -U/2 & t \\
 0 &  0 & 0 & 0 & t & -U/2 
 \end{pmatrix},
 \end{equation}
 when we compute the particle part of the spectral function.
 Because computation for the hole part can be performed for the same six-dimensional Hamiltonian on $\mathcal{H}_{(1,1)} \cup \mathcal{H}_{(0,1)}$ thanks to the particle-hole symmetry system, we describe only the particle part in the following explanation of the experiment.
 Moreover, only one of the two eigenstates in the particle sector has non-zero transition amplitude $|\braket{E_n^{(p)} |\tilde{c}_{k\uparrow}^\dag|\mathrm{GS}}|^2$, so we focus on it and call it the excited state $\ket{ES}$.



\begin{figure*}
    \centering
    \includegraphics[scale = 0.3]{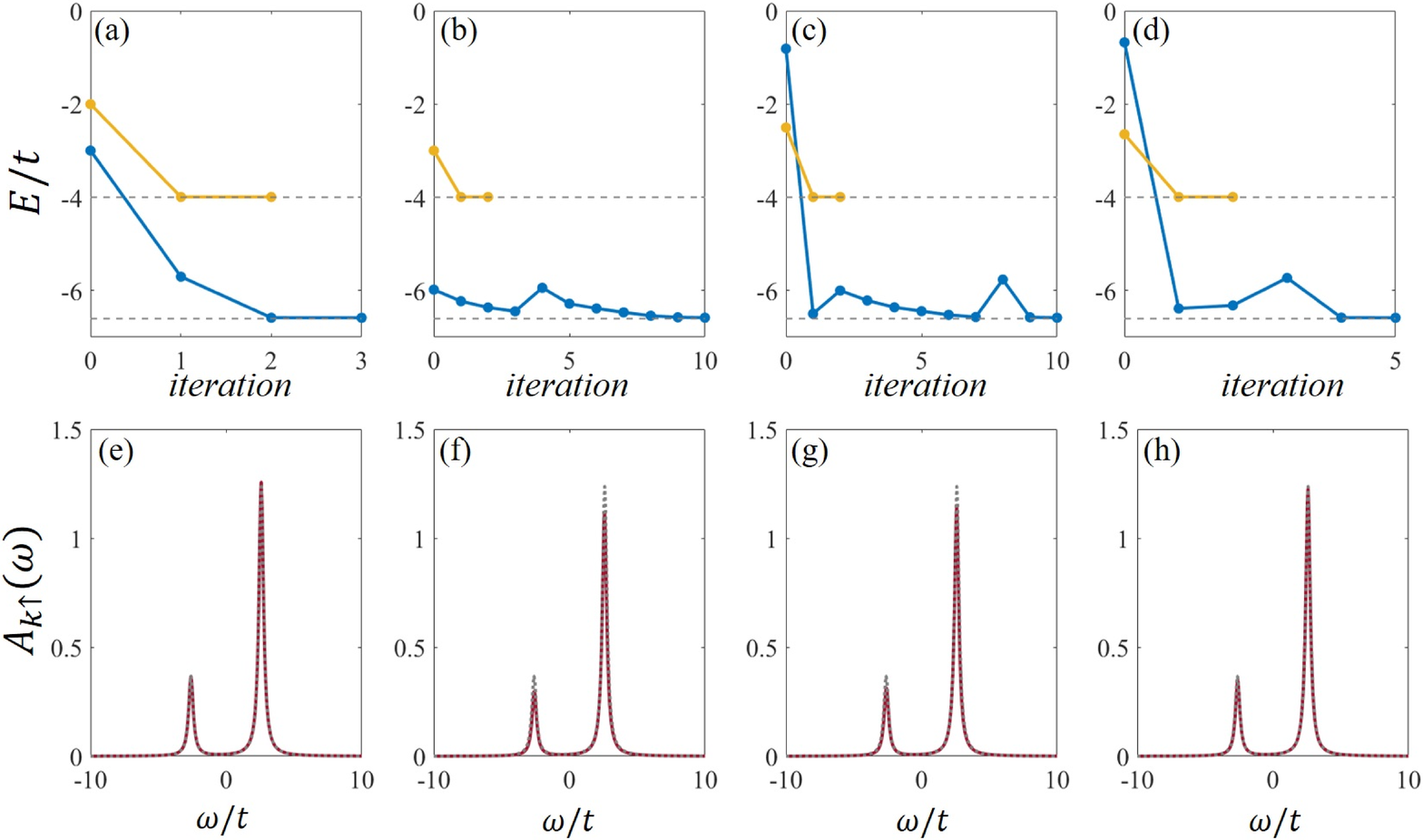}
    \caption{\textbf{Experimental data.} The iteration process is presented in the upper panel with different initial values, $\{\theta_2, \theta_4,\theta_5,\theta_6 \} =$ \textbf{(a)} $\{\pi/2, \pi/2, \pi/2,\pi/2\}$, \textbf{(b)} $\{\pi, \pi, \pi,\pi\}$, \textbf{(c)} $\{0.82,0.99,0.11,0.52\}$ and \textbf{(d)} $\{0.97,0.65,0.91,0.36\}$, respectively. After less than 10 iterations, the measured energies converge to the values closed to the exact values that are indicated by the gray dashed lines. The blue (yellow) lines and dots denote the iteration process of the ground state (excited state energy). The errorbars are too small to be presented in the figures. The final values after convergence are $E_{GS}/t=\{-6.587\pm0.030,~-6.585\pm0.017,~-6.586\pm0.013,~-6.587\pm0.017\}$ and $E_{ES}/t=\{-3.99\pm0.01,~-3.99\pm0.02,~-3.99\pm0.02,~-3.99\pm0.01\}$ corresponding to the four different initial values, respectively. The final values of four parameters are $\{\theta_2,\theta_4,\theta_5,\theta_6\}=\{-1.59,-2.55,-2.56,-1.57\},~\{-1.46,-2.59,-2.52,-1.57\},~\{-1.49,-2.58,-2.53,-1.57\} $ and $\{-1.56,-2.56,-2.55,-1.57\}$. The detailed information of the ground state and excited state can be used to measure the transition amplitude so that the spectral function of the Green's function is obtained and the results are presented in the lower panel, that are shown by the red solid line. They agree well with the numerical exact ones (gray dash line). The figures of \textbf{(e), (f), (g), (h)} correspond to the four initial values respectively. The $\omega$ ranges from $-10$ to $10$.}
    \label{data}
\end{figure*}

{\it Experimental realization.}
We now describe the details of the experimental setup to calculate the Green's function of two-site Fermion Hubbard model. As shown in Fig. \ref{setupfigure}, a 404 nm semiconductor laser illuminates a joint $\beta$-barium-borate (BBO) crystal where the type-I phase-matching spontaneous parametric down conversion (SPDC) process generates the pairs of single photons to serve as the heralded single-photon source. One photon of the pair is sent to the computing module via an optical fiber, whose polarization state of the photon can be adjusted via the half-wave plate (HWP) before BBO, with the other acting as a trigger. The coincidence counter records the coincidence events between two single-photon detectors (SPDs) within 10 s integral time. 

The core of the experimental setup is the computing module which contains the ansatz state preparation part and the measurement part.
In the ansatz state preparation part, we use the the polarization ($(H\rangle, |V\rangle$) and path $(|\mr{path}0\rangle, |\mr{path}1\rangle$, and $|\mr{path}2\rangle)$ degrees of freedom of photon to prepare the six-dimensional ansatz states with the basis $\{|k\rangle\}_{k=1}^6 = \{|H, \mr{path}0\rangle, | V, \mr{path}0\rangle, |H, \mr{path}1\rangle, |V, \mr{path}1\rangle, |H, \mr{path}2\rangle, \\|V, \mr{path}2\rangle\}$.
Here, $\ket{H} (\ket{V})$ represents the horizontal (vertical) polarization of the photon.
The six-dimensional ansatz state of the photon is realized by two beam displacers (BDs) and five rotational operators $R_Y(\theta) = {\cos}(\theta/2)(|H\rangle\langle H| + |V\rangle\langle V|) + {\sin}(\theta/2)(|H\rangle\langle V| - |V\rangle \langle H|)$.
The BDs make $|H\rangle$ beam shift up spatially and keep $|V\rangle$ in the original path.
The rotational operators are constituted by two half-wave plates (HWPs) set at $\theta/4$ and $0^\circ$.
We take an input state for the ansatz preparation part as $|\psi_0\rangle = |H, \mr{path}2\rangle$.
After traveling the two BDs and the five $R_Y(\theta_i)$s and one auxiliary $R_Y$, the ansatz state is parameterized by five parameters $\bm{\theta} = (\theta_1,\ldots,\theta_5)$ and written as
\begin{equation} \label{eq: ansatz state}
    \ket{\psi(\bm{\theta})} = \cos \frac{\theta_1}{2} \ket{\psi_{GS}(\theta_2,\theta_4,\theta_5)} - \sin \frac{\theta_1}{2} \ket{\psi_{ES}(\theta_6)},
\end{equation}
where 
\begin{equation} \label{gsandesansatz}
\begin{split}
|\psi_{GS}\rangle  &= {\cos}(\theta_2/2){\cos}(\theta_4/2)|1\rangle - {\cos}(\theta_2/2){\sin}(\theta_4/2)|2\rangle\\
&-{\sin}(\theta_2/2){\sin}(\theta_5/2)|3\rangle -{\sin}(\theta_2/2){\cos}(\theta_5/2)|4\rangle,\\
|\psi_{ES}\rangle & = {\sin}(\theta_6/2)|5\rangle + {\cos}(\theta_6/2)|6\rangle.
\end{split}
\end{equation}
Here, the component of the ansatz state residing in the Hilbert space corresponding to $\mathcal{H}_{(0,1)} (\mathcal{H}_{(2,1)})$ is explicitly written as $\ket{\psi_{GS}} (\ket{\psi_{ES}})$.

The measurement part evaluates expectation values of observables in the six-dimensional space.
It consists of a polarization beam splitter (PBS), two BDs and six motor rotating plates (MRPs) that are controlled by the computer.
The MRPs are carefully set so that we can calculate an expectation value of any given observables.
For example, when we measure the expectation value of the Hamiltonian, 
we decompose the Hamiltonian into three terms, $H_\mr{six} = tZ\otimes (\ket{\mr{path}0}\bra{\mr{path}1}+\ket{\mr{path}1}\bra{\mr{path}0}) + tX\otimes (-\ket{\mr{path}0}\bra{\mr{path}0}+\ket{\mr{path}1}\bra{\mr{path}1}+\ket{\mr{path}2}\bra{\mr{path}2}) + (\mr{diagonal~terms~in~}\ket{k})$,  where $X,Z$ are Pauli operators for the polarization degrees of freedom $\{\ket{H},\ket{V}\}$. 
The expectation value of the last term is easy to evaluate because it is diagonal in $\ket{k}$ and we simply count the probability of photon in each state of $\ket{k}$ to obtain the expectation value.
The expectation values of the other terms are measured by a similar way with settings of MRPs so as to make the terms diagonal in $\ket{k}$.


To compute the spectral function (Eq.~\eqref{eq: spectral function}) by using this photonic system, we follow the method proposed in Ref.~\cite{Endo2020calculation}.
We first search the ground state $\ket{GS}$ and the excited state $\ket{ES}$ of the system~\eqref{eq: six-dim Ham} by using the variational quantum eigensolver (VQE)~\cite{Peruzzo2014}.
We fix $\theta_1 = 0 \: (-\pi)$ to find the ground (excited) state in the search according to Eq.~\eqref{eq: ansatz state}.
The expectation value $\braket{\psi(\bm{\theta})|H_\mr{six}|\psi(\bm{\theta})}$ is measured by the computational module and the parameters $\bm{\theta}$ are updated to minimize the expectation value by the method proposed in Ref.~\cite{nakanishi2020}, which leverages the trigonometric form of the expectation value with respect to $\bm{\theta}$.
This optimization method is gradient-free and robust to the noise in the experiment.
The obtained ground (excited) state can be written as $\ket{\psi_{GS}(\theta_2^*,\theta_4^*, \theta_5^*)}$ ($\ket{\psi_{ES}(\theta_6^*)}$).
Then, the transition amplitude $|\braket{\psi_{ES}(\theta_6^*)|c_{k\uparrow}^\dag|\psi_{GS}(\theta_2^*,\theta_4^*,\theta_5^*)}|^2$ is measured by evaluating the expectation value of $ c_{k\uparrow}|\psi_{ES}(\theta_6^*)\rangle\langle \psi_{ES}(\theta_6^*)|c_{k\uparrow}^\dag$ of $\ket{\psi_{GS}(\theta_2^*,\theta_4^*, \theta_5^*)}$. 


In experiment, we set $U/t=6$ in the model~\eqref{eq: model}.
We choose four different initial values of $\{\theta_2, \theta_4,\theta_5,\theta_6 \}$, i.e.,  $\{\pi/2, \pi/2, \pi/2,\pi/2\}$, $\{\pi, \pi, \pi,\pi\}$, $\{0.82,0.99,0.11,0.52\}$ and $\{0.97,0.65,0.91,0.36\}$ (the last two are random initial values).
We set the convergence criteria for updating $\bm{\theta}$ in that the relative error between the new expectation value and the old one becomes less than $10^{-3}$.
As shown in Fig.~\ref{data}, the iteration is completed in less than 10 steps, especially only 3 steps for excited states.
After the optimization of $\bm{\theta}$, the obtained grand state energy is $E_{GS}/t=\{-6.587\pm0.030,~-6.585\pm0.017,~-6.586\pm0.013,~-6.587\pm0.017\}$, corresponding to the four different initial values, and the excited state energy is $E_{ES}/t=\{-3.99\pm0.01,~-3.99\pm0.02,~-3.99\pm0.02,~-3.99\pm0.01\}$.
The error bars come from the Poissonian distribution of the photon counts.
The exact values of the ground state energy and the excited state energy are $-6.6056$ and $-4.0$, respectively, as the gray dashed line shown in Fig. \ref{data}.
The obtained energy by the experiment agrees with the exact results within the error bars though some systematic deviation can be seen in the Fig.~\ref{data}.
Then the transition amplitude $|\braket{ES|c_{k\uparrow}|GS}|^2$ is evaluated.
Final results of the spectral function including the contributions both from the particle and hole parts are presented in the lower panels of Fig.~\ref{data}.
The results agrees well with the numerical exact ones.

{\it Discussion and Conclusion.}
In summary, we use the photonic system as a platform to calculate the Green's function of the two-site Fermion Hubbard model~\eqref{eq: model}.
By utilizing the polarization and the path degrees of freedoms, we construct the variational quantum state in the six-dimensional Hilbert space. 
The ground state and the excited state are obtained by using the quantum-classical hybrid algorithm~\cite{Peruzzo2014} with the small number of iterations for updating the variational parameters.
The transition amplitudes between the ground state and excited state are then measured based on the obtained ground and excited states.
The spectral function of the Green's function at zero temperature~\eqref{eq: spectral function} is calculated according to the experimental results and it exhibits almost perfect agreements with the exact results.
The results presented in this letter opens a new door to solving the problems in quantum many-body systems, especially for strongly-correlated systems.

In Refs.~\cite{rungger2020dynamical, Keen_2020}, the Green's function of a four-spin model was computed on the superconducting qubit and the ion-trap devices by a similar way of ours.
Compared to them, our photonic system is much less noisy and we obtain the experimental result very close to the exact values without performing any error mitigation technique~\cite{Temme2017,Endo2018,Songeaaw2019,Kandala2019}.
This illustrates the advantage and possibility of the photonic system as a platform for quantum simulations.
The system and techniques used in this work can be applied to other problems that require larger Hilbert spaces.
The path degree of freedom of the photon can be extended to higher dimension (i.e., more paths) without any technical difficulties.
Moreover, as mentioned in the introduction, the orbital angular momentum or hybrid degrees of freedom of photons can be adopted to greatly extend the Hilbert space of the system with the development of integrated optical technology~\cite{arrazola2021,qiang2021,wang2020}.
This paves a new way to solve larger and more complicated problems.

 Here we combine an intermediate-scale quantum device with classical computation to realize a hybrid algorithm, similarly to variational quantum eigensolver (VQE)~\cite{Peruzzo2014} and variational quantum simulation (VQS)~\cite{ying2017,yuan2019theory} to solve many-body problems. It should be noted that photonic system is an excellent and promising candidate for constructing low-noise intermediate-scale quantum devices, as shown recently~\cite{Zhong2020}. 
 Our work exemplifies such possibility for the photonic system and also inspires calculations for more complicated quantities in quantum mechanics such as the out-of-time-order correlators (OTOC)~\cite{larkin1969quasiclassical,jun2017nmr}.

{\it Acknowledgements---} Jie Zhu thanks Chao Zhang for helpful discussions. This work is funded by the National Natural Science Foundation of China (Grants Nos.~11674306 and 92065113) and Anhui Initiative in Quantum Information Technologies.

\bibliography{bibliography}

\begin{thebibliography}{45}
\expandafter\ifx\csname natexlab\endcsname\relax\def\natexlab#1{#1}\fi
\expandafter\ifx\csname bibnamefont\endcsname\relax
  \def\bibnamefont#1{#1}\fi
\expandafter\ifx\csname bibfnamefont\endcsname\relax
  \def\bibfnamefont#1{#1}\fi
\expandafter\ifx\csname citenamefont\endcsname\relax
  \def\citenamefont#1{#1}\fi
\expandafter\ifx\csname url\endcsname\relax
  \def\url#1{\texttt{#1}}\fi
\expandafter\ifx\csname urlprefix\endcsname\relax\def\urlprefix{URL }\fi
\providecommand{\bibinfo}[2]{#2}
\providecommand{\eprint}[2][]{\url{#2}}

\bibitem[{\citenamefont{Dagotto}(2005)}]{Dagotto2015}
\bibinfo{author}{\bibfnamefont{E.}~\bibnamefont{Dagotto}},
  \bibinfo{journal}{Science} \textbf{\bibinfo{volume}{309}},
  \bibinfo{pages}{257} (\bibinfo{year}{2005}).

\bibitem[{\citenamefont{Bednorz and M{\"u}ller}(1986)}]{Bednorz1986}
\bibinfo{author}{\bibfnamefont{J.~G.} \bibnamefont{Bednorz}} \bibnamefont{and}
  \bibinfo{author}{\bibfnamefont{K.~A.} \bibnamefont{M{\"u}ller}},
  \bibinfo{journal}{Zeitschrift f{\"u}r Physik B Condensed Matter}
  \textbf{\bibinfo{volume}{64}}, \bibinfo{pages}{189} (\bibinfo{year}{1986}).

\bibitem[{\citenamefont{Bonch-Bruevich and Tyablikov}(2015)}]{bonch2015green}
\bibinfo{author}{\bibfnamefont{V.~L.} \bibnamefont{Bonch-Bruevich}}
  \bibnamefont{and} \bibinfo{author}{\bibfnamefont{S.~V.}
  \bibnamefont{Tyablikov}}, \emph{\bibinfo{title}{The Green function method in
  statistical mechanics}} (\bibinfo{publisher}{Courier Dover Publications},
  \bibinfo{year}{2015}).

\bibitem[{\citenamefont{Abrikosov et~al.}(2012)\citenamefont{Abrikosov, Gorkov,
  and Dzyaloshinski}}]{abrikosov2012methods}
\bibinfo{author}{\bibfnamefont{A.~A.} \bibnamefont{Abrikosov}},
  \bibinfo{author}{\bibfnamefont{L.~P.} \bibnamefont{Gorkov}},
  \bibnamefont{and} \bibinfo{author}{\bibfnamefont{I.~E.}
  \bibnamefont{Dzyaloshinski}}, \emph{\bibinfo{title}{Methods of quantum field
  theory in statistical physics}} (\bibinfo{publisher}{Courier Corporation},
  \bibinfo{year}{2012}).

\bibitem[{\citenamefont{Fetter and Walecka}(2012)}]{fetter2012quantum}
\bibinfo{author}{\bibfnamefont{A.~L.} \bibnamefont{Fetter}} \bibnamefont{and}
  \bibinfo{author}{\bibfnamefont{J.~D.} \bibnamefont{Walecka}},
  \emph{\bibinfo{title}{Quantum theory of many-particle systems}}
  (\bibinfo{publisher}{Courier Corporation}, \bibinfo{year}{2012}).

\bibitem[{\citenamefont{Coey}(2010)}]{coey2010magnetism}
\bibinfo{author}{\bibfnamefont{J.~M.} \bibnamefont{Coey}},
  \emph{\bibinfo{title}{Magnetism and magnetic materials}}
  (\bibinfo{publisher}{Cambridge University Press}, \bibinfo{year}{2010}).

\bibitem[{\citenamefont{Hasan and Kane}(2010)}]{hasan2010colloquium}
\bibinfo{author}{\bibfnamefont{M.~Z.} \bibnamefont{Hasan}} \bibnamefont{and}
  \bibinfo{author}{\bibfnamefont{C.~L.} \bibnamefont{Kane}},
  \bibinfo{journal}{Rev. Mod. Phys.} \textbf{\bibinfo{volume}{82}},
  \bibinfo{pages}{3045} (\bibinfo{year}{2010}).

\bibitem[{\citenamefont{Kitaev}(1995)}]{Kitaev1995}
\bibinfo{author}{\bibfnamefont{A.~Y.} \bibnamefont{Kitaev}},
  \bibinfo{journal}{arXiv: 9511026}  (\bibinfo{year}{1995}).

\bibitem[{\citenamefont{Cleve et~al.}(1998)\citenamefont{Cleve, Ekert,
  Macchiavello, and Mosca}}]{Cleve1998}
\bibinfo{author}{\bibfnamefont{R.}~\bibnamefont{Cleve}},
  \bibinfo{author}{\bibfnamefont{A.}~\bibnamefont{Ekert}},
  \bibinfo{author}{\bibfnamefont{C.}~\bibnamefont{Macchiavello}},
  \bibnamefont{and} \bibinfo{author}{\bibfnamefont{M.}~\bibnamefont{Mosca}},
  \bibinfo{journal}{Proceedings of the Royal Society of London. Series A:
  Mathematical, Physical and Engineering Sciences}
  \textbf{\bibinfo{volume}{454}}, \bibinfo{pages}{339} (\bibinfo{year}{1998}).

\bibitem[{\citenamefont{Shor}(1997)}]{shor1997}
\bibinfo{author}{\bibfnamefont{P.~W.} \bibnamefont{Shor}},
  \bibinfo{journal}{SIAM Journal on Computing} \textbf{\bibinfo{volume}{26}},
  \bibinfo{pages}{1484} (\bibinfo{year}{1997}).

\bibitem[{\citenamefont{Preskill}(2018)}]{Preskill2018}
\bibinfo{author}{\bibfnamefont{J.}~\bibnamefont{Preskill}},
  \bibinfo{journal}{Quantum} \textbf{\bibinfo{volume}{2}}, \bibinfo{pages}{79}
  (\bibinfo{year}{2018}).

\bibitem[{\citenamefont{Peruzzo et~al.}(2014)\citenamefont{Peruzzo, McClean,
  Shadbolt, Yung, Zhou, Love, Aspuru-Guzik, and O'brien}}]{Peruzzo2014}
\bibinfo{author}{\bibfnamefont{A.}~\bibnamefont{Peruzzo}},
  \bibinfo{author}{\bibfnamefont{J.}~\bibnamefont{McClean}},
  \bibinfo{author}{\bibfnamefont{P.}~\bibnamefont{Shadbolt}},
  \bibinfo{author}{\bibfnamefont{M.-H.} \bibnamefont{Yung}},
  \bibinfo{author}{\bibfnamefont{X.-Q.} \bibnamefont{Zhou}},
  \bibinfo{author}{\bibfnamefont{P.~J.} \bibnamefont{Love}},
  \bibinfo{author}{\bibfnamefont{A.}~\bibnamefont{Aspuru-Guzik}},
  \bibnamefont{and} \bibinfo{author}{\bibfnamefont{J.~L.}
  \bibnamefont{O'brien}}, \bibinfo{journal}{Nat. Commun.}
  \textbf{\bibinfo{volume}{5}}, \bibinfo{pages}{4213} (\bibinfo{year}{2014}).

\bibitem[{\citenamefont{Cerezo et~al.}(2020)}]{cerezo2020variational}
\bibinfo{author}{\bibfnamefont{M.}~\bibnamefont{Cerezo}} \bibnamefont{et~al.},
  \bibinfo{journal}{arXiv:2012.09265}  (\bibinfo{year}{2020}).

\bibitem[{\citenamefont{Endo et~al.}(2021)\citenamefont{Endo, Cai, Benjamin,
  and Yuan}}]{Endo2021review}
\bibinfo{author}{\bibfnamefont{S.}~\bibnamefont{Endo}},
  \bibinfo{author}{\bibfnamefont{Z.}~\bibnamefont{Cai}},
  \bibinfo{author}{\bibfnamefont{S.~C.} \bibnamefont{Benjamin}},
  \bibnamefont{and} \bibinfo{author}{\bibfnamefont{X.}~\bibnamefont{Yuan}},
  \bibinfo{journal}{Journal of the Physical Society of Japan}
  \textbf{\bibinfo{volume}{90}}, \bibinfo{pages}{032001}
  (\bibinfo{year}{2021}).

\bibitem[{\citenamefont{Kok et~al.}(2007)\citenamefont{Kok, Munro, Nemoto,
  Ralph, Dowling, and Milburn}}]{pieter2007}
\bibinfo{author}{\bibfnamefont{P.}~\bibnamefont{Kok}},
  \bibinfo{author}{\bibfnamefont{W.~J.} \bibnamefont{Munro}},
  \bibinfo{author}{\bibfnamefont{K.}~\bibnamefont{Nemoto}},
  \bibinfo{author}{\bibfnamefont{T.~C.} \bibnamefont{Ralph}},
  \bibinfo{author}{\bibfnamefont{J.~P.} \bibnamefont{Dowling}},
  \bibnamefont{and} \bibinfo{author}{\bibfnamefont{G.~J.}
  \bibnamefont{Milburn}}, \bibinfo{journal}{Rev. Mod. Phys.}
  \textbf{\bibinfo{volume}{79}}, \bibinfo{pages}{135} (\bibinfo{year}{2007}).

\bibitem[{\citenamefont{Aspuru-Guzik and Walther}(2012)}]{aspuru2012}
\bibinfo{author}{\bibfnamefont{A.}~\bibnamefont{Aspuru-Guzik}}
  \bibnamefont{and} \bibinfo{author}{\bibfnamefont{P.}~\bibnamefont{Walther}},
  \bibinfo{journal}{Nat. Phys.} \textbf{\bibinfo{volume}{8}},
  \bibinfo{pages}{285} (\bibinfo{year}{2012}).

\bibitem[{\citenamefont{Poh et~al.}(2015)\citenamefont{Poh, Joshi, Cer\`e,
  Cabello, and Kurtsiefer}}]{poh2015}
\bibinfo{author}{\bibfnamefont{H.~S.} \bibnamefont{Poh}},
  \bibinfo{author}{\bibfnamefont{S.~K.} \bibnamefont{Joshi}},
  \bibinfo{author}{\bibfnamefont{A.}~\bibnamefont{Cer\`e}},
  \bibinfo{author}{\bibfnamefont{A.}~\bibnamefont{Cabello}}, \bibnamefont{and}
  \bibinfo{author}{\bibfnamefont{C.}~\bibnamefont{Kurtsiefer}},
  \bibinfo{journal}{Phys. Rev. Lett.} \textbf{\bibinfo{volume}{115}},
  \bibinfo{pages}{180408} (\bibinfo{year}{2015}).

\bibitem[{\citenamefont{Hu et~al.}(2020)\citenamefont{Hu, Xing, Liu, Huang, Li,
  Guo, Erker, and Huber}}]{xiaominhu2020dim32}
\bibinfo{author}{\bibfnamefont{X.-M.} \bibnamefont{Hu}},
  \bibinfo{author}{\bibfnamefont{W.-B.} \bibnamefont{Xing}},
  \bibinfo{author}{\bibfnamefont{B.-H.} \bibnamefont{Liu}},
  \bibinfo{author}{\bibfnamefont{Y.-F.} \bibnamefont{Huang}},
  \bibinfo{author}{\bibfnamefont{C.-F.} \bibnamefont{Li}},
  \bibinfo{author}{\bibfnamefont{G.-C.} \bibnamefont{Guo}},
  \bibinfo{author}{\bibfnamefont{P.}~\bibnamefont{Erker}}, \bibnamefont{and}
  \bibinfo{author}{\bibfnamefont{M.}~\bibnamefont{Huber}},
  \bibinfo{journal}{Phys. Rev. Lett.} \textbf{\bibinfo{volume}{125}},
  \bibinfo{pages}{090503} (\bibinfo{year}{2020}).

\bibitem[{\citenamefont{Fickler et~al.}(2016)\citenamefont{Fickler, Campbell,
  Buchler, Lam, and Zeilinger}}]{oam10010}
\bibinfo{author}{\bibfnamefont{R.}~\bibnamefont{Fickler}},
  \bibinfo{author}{\bibfnamefont{G.}~\bibnamefont{Campbell}},
  \bibinfo{author}{\bibfnamefont{B.}~\bibnamefont{Buchler}},
  \bibinfo{author}{\bibfnamefont{P.~K.} \bibnamefont{Lam}}, \bibnamefont{and}
  \bibinfo{author}{\bibfnamefont{A.}~\bibnamefont{Zeilinger}},
  \bibinfo{journal}{Proceedings of the National Academy of Sciences}
  \textbf{\bibinfo{volume}{113}}, \bibinfo{pages}{13642}
  (\bibinfo{year}{2016}).

\bibitem[{\citenamefont{Wang et~al.}(2018)}]{xilinwang2018}
\bibinfo{author}{\bibfnamefont{X.-L.} \bibnamefont{Wang}} \bibnamefont{et~al.},
  \bibinfo{journal}{Phys. Rev. Lett.} \textbf{\bibinfo{volume}{120}},
  \bibinfo{pages}{260502} (\bibinfo{year}{2018}).

\bibitem[{\citenamefont{Endo et~al.}(2020)\citenamefont{Endo, Kurata, and
  Nakagawa}}]{Endo2020calculation}
\bibinfo{author}{\bibfnamefont{S.}~\bibnamefont{Endo}},
  \bibinfo{author}{\bibfnamefont{I.}~\bibnamefont{Kurata}}, \bibnamefont{and}
  \bibinfo{author}{\bibfnamefont{Y.~O.} \bibnamefont{Nakagawa}},
  \bibinfo{journal}{Phys. Rev. Research} \textbf{\bibinfo{volume}{2}},
  \bibinfo{pages}{033281} (\bibinfo{year}{2020}).

\bibitem[{\citenamefont{Bauer et~al.}(2016)\citenamefont{Bauer, Wecker, Millis,
  Hastings, and Troyer}}]{Bauer2016}
\bibinfo{author}{\bibfnamefont{B.}~\bibnamefont{Bauer}},
  \bibinfo{author}{\bibfnamefont{D.}~\bibnamefont{Wecker}},
  \bibinfo{author}{\bibfnamefont{A.~J.} \bibnamefont{Millis}},
  \bibinfo{author}{\bibfnamefont{M.~B.} \bibnamefont{Hastings}},
  \bibnamefont{and} \bibinfo{author}{\bibfnamefont{M.}~\bibnamefont{Troyer}},
  \bibinfo{journal}{Phys. Rev. X} \textbf{\bibinfo{volume}{6}},
  \bibinfo{pages}{031045} (\bibinfo{year}{2016}).

\bibitem[{\citenamefont{Kreula et~al.}(2016)\citenamefont{Kreula, Clark, and
  Jaksch}}]{Kreula2016}
\bibinfo{author}{\bibfnamefont{J.}~\bibnamefont{Kreula}},
  \bibinfo{author}{\bibfnamefont{S.~R.} \bibnamefont{Clark}}, \bibnamefont{and}
  \bibinfo{author}{\bibfnamefont{D.}~\bibnamefont{Jaksch}},
  \bibinfo{journal}{Sci. Rep.} \textbf{\bibinfo{volume}{6}},
  \bibinfo{pages}{32940} (\bibinfo{year}{2016}).

\bibitem[{\citenamefont{Wecker et~al.}(2015)\citenamefont{Wecker, Hastings,
  Wiebe, Clark, Nayak, and Troyer}}]{Wecker2015}
\bibinfo{author}{\bibfnamefont{D.}~\bibnamefont{Wecker}},
  \bibinfo{author}{\bibfnamefont{M.~B.} \bibnamefont{Hastings}},
  \bibinfo{author}{\bibfnamefont{N.}~\bibnamefont{Wiebe}},
  \bibinfo{author}{\bibfnamefont{B.~K.} \bibnamefont{Clark}},
  \bibinfo{author}{\bibfnamefont{C.}~\bibnamefont{Nayak}}, \bibnamefont{and}
  \bibinfo{author}{\bibfnamefont{M.}~\bibnamefont{Troyer}},
  \bibinfo{journal}{Phys. Rev. A} \textbf{\bibinfo{volume}{92}},
  \bibinfo{pages}{062318} (\bibinfo{year}{2015}).

\bibitem[{\citenamefont{Kosugi and Matsushita}(2020)}]{Kosugi2020}
\bibinfo{author}{\bibfnamefont{T.}~\bibnamefont{Kosugi}} \bibnamefont{and}
  \bibinfo{author}{\bibfnamefont{Y.}~\bibnamefont{Matsushita}},
  \bibinfo{journal}{Phys. Rev. A} \textbf{\bibinfo{volume}{101}},
  \bibinfo{pages}{012330} (\bibinfo{year}{2020}).

\bibitem[{\citenamefont{Pedernales et~al.}(2014)\citenamefont{Pedernales,
  Di~Candia, Egusquiza, Casanova, and Solano}}]{Pedernales2014}
\bibinfo{author}{\bibfnamefont{J.~S.} \bibnamefont{Pedernales}},
  \bibinfo{author}{\bibfnamefont{R.}~\bibnamefont{Di~Candia}},
  \bibinfo{author}{\bibfnamefont{I.~L.} \bibnamefont{Egusquiza}},
  \bibinfo{author}{\bibfnamefont{J.}~\bibnamefont{Casanova}}, \bibnamefont{and}
  \bibinfo{author}{\bibfnamefont{E.}~\bibnamefont{Solano}},
  \bibinfo{journal}{Phys. Rev. Lett.} \textbf{\bibinfo{volume}{113}},
  \bibinfo{pages}{020505} (\bibinfo{year}{2014}).

\bibitem[{\citenamefont{Roggero and Carlson}(2019)}]{Roggero2019}
\bibinfo{author}{\bibfnamefont{A.}~\bibnamefont{Roggero}} \bibnamefont{and}
  \bibinfo{author}{\bibfnamefont{J.}~\bibnamefont{Carlson}},
  \bibinfo{journal}{Phys. Rev. C} \textbf{\bibinfo{volume}{100}},
  \bibinfo{pages}{034610} (\bibinfo{year}{2019}).

\bibitem[{\citenamefont{Gutzwiller}(1963)}]{Gutzwiller1963}
\bibinfo{author}{\bibfnamefont{M.~C.} \bibnamefont{Gutzwiller}},
  \bibinfo{journal}{Phys. Rev. Lett.} \textbf{\bibinfo{volume}{10}},
  \bibinfo{pages}{159} (\bibinfo{year}{1963}).

\bibitem[{\citenamefont{Hubbard}(1963)}]{Hubbard1963}
\bibinfo{author}{\bibfnamefont{J.}~\bibnamefont{Hubbard}},
  \bibinfo{journal}{Proceedings of the Royal Society of London. Series A.
  Mathematical and Physical Sciences} \textbf{\bibinfo{volume}{276}},
  \bibinfo{pages}{238} (\bibinfo{year}{1963}).

\bibitem[{\citenamefont{Kanamori}(1963)}]{Kanamori1963}
\bibinfo{author}{\bibfnamefont{J.}~\bibnamefont{Kanamori}},
  \bibinfo{journal}{Progress of Theoretical Physics}
  \textbf{\bibinfo{volume}{30}}, \bibinfo{pages}{275} (\bibinfo{year}{1963}).

\bibitem[{\citenamefont{Nakanishi et~al.}(2020)\citenamefont{Nakanishi, Fujii,
  and Todo}}]{nakanishi2020}
\bibinfo{author}{\bibfnamefont{K.~M.} \bibnamefont{Nakanishi}},
  \bibinfo{author}{\bibfnamefont{K.}~\bibnamefont{Fujii}}, \bibnamefont{and}
  \bibinfo{author}{\bibfnamefont{S.}~\bibnamefont{Todo}},
  \bibinfo{journal}{Phys. Rev. Research} \textbf{\bibinfo{volume}{2}},
  \bibinfo{pages}{043158} (\bibinfo{year}{2020}).

\bibitem[{\citenamefont{Rungger et~al.}(2020)\citenamefont{Rungger,
  Fitzpatrick, Chen, Alderete, Apel, Cowtan, Patterson, Ramo, Zhu, Nguyen
  et~al.}}]{rungger2020dynamical}
\bibinfo{author}{\bibfnamefont{I.}~\bibnamefont{Rungger}},
  \bibinfo{author}{\bibfnamefont{N.}~\bibnamefont{Fitzpatrick}},
  \bibinfo{author}{\bibfnamefont{H.}~\bibnamefont{Chen}},
  \bibinfo{author}{\bibfnamefont{C.~H.} \bibnamefont{Alderete}},
  \bibinfo{author}{\bibfnamefont{H.}~\bibnamefont{Apel}},
  \bibinfo{author}{\bibfnamefont{A.}~\bibnamefont{Cowtan}},
  \bibinfo{author}{\bibfnamefont{A.}~\bibnamefont{Patterson}},
  \bibinfo{author}{\bibfnamefont{D.~M.} \bibnamefont{Ramo}},
  \bibinfo{author}{\bibfnamefont{Y.}~\bibnamefont{Zhu}},
  \bibinfo{author}{\bibfnamefont{N.~H.} \bibnamefont{Nguyen}},
  \bibnamefont{et~al.}, \bibinfo{journal}{arXiv:1910.04735}
  (\bibinfo{year}{2020}).

\bibitem[{\citenamefont{Keen et~al.}(2020)\citenamefont{Keen, Maier, Johnston,
  and Lougovski}}]{Keen_2020}
\bibinfo{author}{\bibfnamefont{T.}~\bibnamefont{Keen}},
  \bibinfo{author}{\bibfnamefont{T.}~\bibnamefont{Maier}},
  \bibinfo{author}{\bibfnamefont{S.}~\bibnamefont{Johnston}}, \bibnamefont{and}
  \bibinfo{author}{\bibfnamefont{P.}~\bibnamefont{Lougovski}},
  \bibinfo{journal}{Quantum Science and Technology}
  \textbf{\bibinfo{volume}{5}}, \bibinfo{pages}{035001} (\bibinfo{year}{2020}).

\bibitem[{\citenamefont{Temme et~al.}(2017)\citenamefont{Temme, Bravyi, and
  Gambetta}}]{Temme2017}
\bibinfo{author}{\bibfnamefont{K.}~\bibnamefont{Temme}},
  \bibinfo{author}{\bibfnamefont{S.}~\bibnamefont{Bravyi}}, \bibnamefont{and}
  \bibinfo{author}{\bibfnamefont{J.~M.} \bibnamefont{Gambetta}},
  \bibinfo{journal}{Phys. Rev. Lett.} \textbf{\bibinfo{volume}{119}},
  \bibinfo{pages}{180509} (\bibinfo{year}{2017}).

\bibitem[{\citenamefont{Endo et~al.}(2018)\citenamefont{Endo, Benjamin, and
  Li}}]{Endo2018}
\bibinfo{author}{\bibfnamefont{S.}~\bibnamefont{Endo}},
  \bibinfo{author}{\bibfnamefont{S.~C.} \bibnamefont{Benjamin}},
  \bibnamefont{and} \bibinfo{author}{\bibfnamefont{Y.}~\bibnamefont{Li}},
  \bibinfo{journal}{Phys. Rev. X} \textbf{\bibinfo{volume}{8}},
  \bibinfo{pages}{031027} (\bibinfo{year}{2018}).

\bibitem[{\citenamefont{Song et~al.}(2019)\citenamefont{Song, Cui, Wang, Hao,
  Feng, and Li}}]{Songeaaw2019}
\bibinfo{author}{\bibfnamefont{C.}~\bibnamefont{Song}},
  \bibinfo{author}{\bibfnamefont{J.}~\bibnamefont{Cui}},
  \bibinfo{author}{\bibfnamefont{H.}~\bibnamefont{Wang}},
  \bibinfo{author}{\bibfnamefont{J.}~\bibnamefont{Hao}},
  \bibinfo{author}{\bibfnamefont{H.}~\bibnamefont{Feng}}, \bibnamefont{and}
  \bibinfo{author}{\bibfnamefont{Y.}~\bibnamefont{Li}}, \bibinfo{journal}{Sci.
  Adv.} \textbf{\bibinfo{volume}{5}} (\bibinfo{year}{2019}).

\bibitem[{\citenamefont{Kandala et~al.}(2019)\citenamefont{Kandala, Temme,
  C{\'o}rcoles, Mezzacapo, Chow, and Gambetta}}]{Kandala2019}
\bibinfo{author}{\bibfnamefont{A.}~\bibnamefont{Kandala}},
  \bibinfo{author}{\bibfnamefont{K.}~\bibnamefont{Temme}},
  \bibinfo{author}{\bibfnamefont{A.~D.} \bibnamefont{C{\'o}rcoles}},
  \bibinfo{author}{\bibfnamefont{A.}~\bibnamefont{Mezzacapo}},
  \bibinfo{author}{\bibfnamefont{J.~M.} \bibnamefont{Chow}}, \bibnamefont{and}
  \bibinfo{author}{\bibfnamefont{J.~M.} \bibnamefont{Gambetta}},
  \bibinfo{journal}{Nature} \textbf{\bibinfo{volume}{567}},
  \bibinfo{pages}{491} (\bibinfo{year}{2019}).

\bibitem[{\citenamefont{Arrazola et~al.}(2021)}]{arrazola2021}
\bibinfo{author}{\bibfnamefont{J.}~\bibnamefont{Arrazola}}
  \bibnamefont{et~al.}, \bibinfo{journal}{Nature}
  \textbf{\bibinfo{volume}{591}}, \bibinfo{pages}{54} (\bibinfo{year}{2021}).

\bibitem[{\citenamefont{Qiang et~al.}(2021)}]{qiang2021}
\bibinfo{author}{\bibfnamefont{X.}~\bibnamefont{Qiang}} \bibnamefont{et~al.},
  \bibinfo{journal}{Sci. Adv.} \textbf{\bibinfo{volume}{7}},
  \bibinfo{pages}{eabb8375} (\bibinfo{year}{2021}).

\bibitem[{\citenamefont{Wang et~al.}(2020)\citenamefont{Wang, Sciarrino, Laing,
  and Thompson}}]{wang2020}
\bibinfo{author}{\bibfnamefont{J.}~\bibnamefont{Wang}},
  \bibinfo{author}{\bibfnamefont{F.}~\bibnamefont{Sciarrino}},
  \bibinfo{author}{\bibfnamefont{A.}~\bibnamefont{Laing}}, \bibnamefont{and}
  \bibinfo{author}{\bibfnamefont{M.~G.} \bibnamefont{Thompson}},
  \bibinfo{journal}{Nat. Photonics} \textbf{\bibinfo{volume}{14}},
  \bibinfo{pages}{273} (\bibinfo{year}{2020}).

\bibitem[{\citenamefont{Li and Benjamin}(2017)}]{ying2017}
\bibinfo{author}{\bibfnamefont{Y.}~\bibnamefont{Li}} \bibnamefont{and}
  \bibinfo{author}{\bibfnamefont{S.~C.} \bibnamefont{Benjamin}},
  \bibinfo{journal}{Phys. Rev. X} \textbf{\bibinfo{volume}{7}},
  \bibinfo{pages}{021050} (\bibinfo{year}{2017}).

\bibitem[{\citenamefont{Yuan et~al.}(2019)\citenamefont{Yuan, Endo, Zhao, Li,
  and Benjamin}}]{yuan2019theory}
\bibinfo{author}{\bibfnamefont{X.}~\bibnamefont{Yuan}},
  \bibinfo{author}{\bibfnamefont{S.}~\bibnamefont{Endo}},
  \bibinfo{author}{\bibfnamefont{Q.}~\bibnamefont{Zhao}},
  \bibinfo{author}{\bibfnamefont{Y.}~\bibnamefont{Li}}, \bibnamefont{and}
  \bibinfo{author}{\bibfnamefont{S.~C.} \bibnamefont{Benjamin}},
  \bibinfo{journal}{Quantum} \textbf{\bibinfo{volume}{3}}, \bibinfo{pages}{191}
  (\bibinfo{year}{2019}).

\bibitem[{\citenamefont{Zhong et~al.}(2020)}]{Zhong2020}
\bibinfo{author}{\bibfnamefont{H.-S.} \bibnamefont{Zhong}}
  \bibnamefont{et~al.}, \bibinfo{journal}{Science}
  \textbf{\bibinfo{volume}{370}}, \bibinfo{pages}{1460} (\bibinfo{year}{2020}).

\bibitem[{\citenamefont{Larkin and
  Ovchinnikov}(1969)}]{larkin1969quasiclassical}
\bibinfo{author}{\bibfnamefont{A.}~\bibnamefont{Larkin}} \bibnamefont{and}
  \bibinfo{author}{\bibfnamefont{Y.~N.} \bibnamefont{Ovchinnikov}},
  \bibinfo{journal}{Sov Phys JETP} \textbf{\bibinfo{volume}{28}},
  \bibinfo{pages}{1200} (\bibinfo{year}{1969}).

\bibitem[{\citenamefont{Li et~al.}(2017)\citenamefont{Li, Fan, Wang, Ye, Zeng,
  Zhai, Peng, and Du}}]{jun2017nmr}
\bibinfo{author}{\bibfnamefont{J.}~\bibnamefont{Li}},
  \bibinfo{author}{\bibfnamefont{R.}~\bibnamefont{Fan}},
  \bibinfo{author}{\bibfnamefont{H.}~\bibnamefont{Wang}},
  \bibinfo{author}{\bibfnamefont{B.}~\bibnamefont{Ye}},
  \bibinfo{author}{\bibfnamefont{B.}~\bibnamefont{Zeng}},
  \bibinfo{author}{\bibfnamefont{H.}~\bibnamefont{Zhai}},
  \bibinfo{author}{\bibfnamefont{X.}~\bibnamefont{Peng}}, \bibnamefont{and}
  \bibinfo{author}{\bibfnamefont{J.}~\bibnamefont{Du}}, \bibinfo{journal}{Phys.
  Rev. X} \textbf{\bibinfo{volume}{7}}, \bibinfo{pages}{031011}
  (\bibinfo{year}{2017}).

\end{thebibliography}


\end{document}